\documentclass[12pt]{article}
\newcommand{\mb}[1]{{\mbox{\boldmath{$#1$}}}}
\begin{document}

\title{The induced representation of the isometry group of the 
Euclidean Taub-NUT space and new spherical harmonics}

\author{Ion I. Cot\u aescu \thanks{E-mail:~~~cota@physics.uvt.ro}\\ 
{\small \it West University of Timi\c soara,}\\
       {\small \it V. P\^ arvan Ave. 4, RO-1900 Timi\c soara, Romania}
\and
Mihai Visinescu \thanks{E-mail:~~~mvisin@theor1.theory.nipne.ro}\\
{\small \it Department of Theoretical Physics,}\\
{\small \it National Institute for Physics and Nuclear Engineering,}\\
{\small \it P.O.Box M.G.-6, Magurele, Bucharest, Romania}}
\date{\today}

\maketitle

\begin{abstract}
It is shown that the $SO(3)$ isometries of the Euclidean Taub-NUT space 
combine a linear three-dimensional representation with one induced by a 
$SO(2)$ subgroup, giving the transformation law of the fourth coordinate 
under rotations. This explains the special form of the angular momentum 
operator on this manifold which leads to a new type of spherical harmonics 
and spinors.

~

Keywords: {\it Euclidean Taub-NUT space, spherical harmonics, 
sphe\-ri\-cal spinors, isometry group, induced representation.}

~

Pacs 04.62.+v

\end{abstract}
\

\newpage

\section{Introduction}

The Euclidean Taub-Newman-Unti-Tamburino (Taub-NUT) metric is involved 
in many modern studies in physics. One of the first examples given of a 
gravitational instanton was the self-dual Taub-NUT solution \cite{Ha}.
Much attention has been paid to the Euclidean Taub-NUT metric since in 
the long-distance limit the relative motion of two monopoles is 
described approximately by its geodesics \cite{Ma, AH}. 
On the other 
hand this metric is just the space part of the metric of the celebrated 
Kaluza-Klein monopole of Gross and Perry \cite{GP} and of Sorkin 
\cite{So}.

The Taub-NUT space is also of interest since beside 
isometries there are hidden symmetries giving rise to conserved  
quantities associated to St\" ackel-Killing tensors \cite{GH}. 
There is a conserved vector, analogous to the 
Runge-Lenz vector of the Kepler type problem, whose existence is rather 
surprising in view of the complexity of the equations of motion  
\cite{GM}-\cite{CFH}. These hidden symmetries are related to the existence 
of special objects arising in this geometry, i.e. four Killing-Yano 
tensors \cite{Ya} generating the St\" ackel-Killing ones 
\cite{GR}, \cite{vH}-\cite{VV2}.

The quantum theory in the Euclidean Taub-NUT background has also interesting 
specific features in the case of the scalar fields \cite{CV1} as well as for 
Dirac fields of  spin $\frac{1}{2}$ fermions \cite{CV2}-\cite{CV4}. 
In both cases there exit large 
algebras of conserved observables \cite{CV5} including the components of the 
angular momentum and three components of the Runge-Lenz operator that 
complete a six-dimensional dynamical algebra \cite{GH,CV4,CV5}. 
Remarkably, the orbital angular momentum has a special unusual 
form that generates new harmonics called $SO(3)\otimes U(1)$-harmonics 
\cite{CV1}. What is the explanation of this fact ?

In our opinion the form of the angular momentum operator is determined by 
the specific type of isometries of the Euclidean Taub-NUT space which combine 
linear representations with induced ones \cite{MAK,W}. The purpose of this 
article is to prove this shoving that the fourth Cartesian (or spherical) 
coordinate transforms under rotations according to a representation of the 
$SO(3)$ isometry group induced by one of its $SO(2)$ subgroups. Thus, after we 
present in the next section the form of the isometry transformations, we 
demonstrate in Sec. 3 that these involve an induced representation for which 
we deduce the explicit closed form of the transformation law. The Sec. 4 is 
devoted to the isometry generators showing how the induced representation 
determines the specific form of the angular momentum operators of the scalar 
or spin half theories. The corresponding new spherical harmonics \cite{CV1} 
as well as the associated Pauli spherical spinors \cite{CV2} are briefly 
presented in the next section. Some technical details are given in an Appendix.  
     
\section{$SO(3)\otimes U(1)$ isometry transformations}

The  Euclidean Taub-NUT manifold $M_4$ is a 4-dimensional Kaluza-Klein space 
which has static charts with the Cartesian coordinates 
$x^{\mu}$ ($\mu, \nu,... = 1,2,3,4$) where  $x^{i}$ ($i,j,...=1,2,3$) are 
the {\em physical} Cartesian space coordinates while $x^4$ is the Cartesian 
extra-coordinate. In the usual three-dimensional vector notations,  
${\bf x}=(x^1,x^2,x^3)$, $r=|{\bf x}|$  and $dl^{2}=d{\bf x}\cdot d{\bf x}$,
its line element reads
\begin{equation}\label{(met)}
ds^2=\frac{1}{V(r)}dl^2 + V(r)[dx^4+A_i({\bf x})dx^i]^2
\end{equation}
where 
\begin{equation}\label{(tn)}
\frac{1}{V}=1+\frac{\mu}{r}\,,\quad A_{1}=-\frac{\mu}{r}\frac{x^{2}}{r+x^{3}}\,,
\quad A_{2}=\frac{\mu}{r}\frac{x^{1}}{r+x^{3}}\,,\quad A_{3}=0\,. 
\end{equation}
The real number $\mu$ is the main parameter of the theory. If one interprets, 
${\bf A}$ as the gauge field of a monopole it 
results  the magnetic field with central symmetry
\begin{equation}
{\bf B}\,=\mu\frac{{\bf x}}{r^3}\,.
\end{equation}

Other important charts are those with spherical coordinates 
$(r, \theta, \phi, \chi)$ where  $r, \theta, \phi$, are commonly related 
to the 
physical Cartesian ones, $x^i$. The apparent singularity at the 
origin is unphysical if $x^4$ is periodic with period $4\pi \mu$ 
\cite{GM}-\cite{CFH}. Consequently the fourth coordinate $\chi$ is 
defined 
such that 
\begin{equation}\label{4cp}
x^{4}=-\mu(\chi +\phi)\,.
\end{equation}
This chart covers the domain where $r>0$ if $\mu>0$ or $r>|\mu|$ if $\mu<0$, 
the angular coordinates $\theta,\,\phi$ cover the sphere $S^{2}$ and 
$\chi\in D_{\chi}=[0,4\pi)$.  The line element in spherical coordinates is 
\begin{equation}\label{metsf}
ds^{2}=\frac{1}{V}(dr^{2}+r^{2}d\theta^{2}+ 
r^{2}\sin^{2}\theta\, d\phi^{2})+\mu^{2}V(d\chi+\cos\theta\, d\phi)^{2}\,,  
\end{equation}
since
\begin{equation}
A_{r}=A_{\theta}=0\,,\quad A_{\phi}=\mu(1-\cos\theta)\,.
\end{equation}

The Euclidean Taub-NUT spaces $M_4$ possess a special type of isometries 
which combines the space transformations with the gauge transformations of the 
gauge field ${\bf A}({\bf x})$. There are $U_4(1)$ transformations 
$x^4\to x^{\prime 4}=x^4 +\lambda$  which leave the metric invariant if 
$\lambda$ is a point-independent real constant. Moreover, if one takes   
$\lambda=\lambda({\bf x})$ an arbitrary function of ${\bf x}$ then these 
become gauge transformations preserving the form of the line element only if 
one requires ${\bf A}$ to transform as 
\begin{equation}
A_i({\bf x})\to A'_i({\bf x})=A_i({\bf x})-\partial_i \lambda({\bf x})\,.
\end{equation}  
Thus it is obvious that $U_4(1)$ is an isometry group playing, in addition, 
the role of the gauge group associated to the gauge field ${\bf A}$. In other 
respects,  this geometry allows an $SO(3)$ symmetry given by usual 
{\em linear} rotations of the physical space coordinates, 
${\bf x}\to {{\bf x}\,}'=R {\bf x}$ with $R\in SO(3)$, and the special 
non-linear transformations of the fourth coordinate, 
\begin{equation}\label{ind}
R\,:\quad x^4\to x^{\prime\, 4}=x^4+h(R,{\bf x})\,,   
\end{equation}
produced by a function $h$ depending on $R$ and ${\bf x}$ which must satisfy
\begin{equation}\label{cond} 
h(R' R, {\bf x})=h(R',R{\bf x})+h(R,{\bf x})\,, \quad
h(Id,{\bf x})=0 
\end{equation} 
where $Id$ is the identity of $SO(3)$. Obviously, this condition guarantees that 
Eq. (\ref{ind}) defines a representation of the $SO(3)$ group. These 
transformations preserve the general form of the line element (\ref{(met)}) if 
${\bf A}$ transforms {\em manifestly} covariant under rotations as a vector 
field, up to a gauge transformation, $V$ being a scalar. In this way one 
obtains a representation of the group $SO(3)\otimes U_4(1)$ whose 
transformations,  
\begin{eqnarray}
{\bf x}&\to& {{\bf x}\,}'=R {\bf x}\label{ec1}\\ 
\left[ R,\lambda({\bf x}) \right]\,:\,\qquad x^4&\to& x^{\prime\, 4}
=x^4+h(R,{\bf x})+\lambda({\bf x})\label{ec2}\\   
{\bf A}({\bf x})&\to& {{\bf A}\,}'({{\bf x}\,}')=R\left\{ {\bf A}({\bf x})
-{\mb \nabla}\,
[h(R,{\bf x})+ \lambda({\bf x})]\right\}\,,\label{ec3}
\end{eqnarray}
produced by any $R\in SO(3)$ and real function $\lambda({\bf x})$,  
combines isometries and gauge transformations. Hereby we can {\em separate} 
the isometries requiring that for point-independent parameters $\lambda$ 
the components of the gauge field remain unchanged, i.e. $A'_i=A_i$. According 
to Eq. (\ref{ec3}), this condition can be written as
\begin{equation}\label{hA}
{\mb \nabla}\, h(R,{\bf x})={\bf A}({\bf x})-R^{-1}{\bf A}(R{\bf x})
\end{equation}   
defining the {\em specific} function $h$ corresponding to the gauge field 
${\bf A}$. Therefore, the isometry transformations $(R,\lambda): x\to x'$  
are three-dimensional rotations and  $x^4$ translations  that 
transform  $x=({\bf x},x^4)$  into  $x'=({{\bf x}\,}',x^{\prime\,4})$ 
according to Eqs. (\ref{ec1}) and (\ref{ec2}) restricted to point-independent 
values of $\lambda$ and function $h$ defined by Eq. (\ref{hA}). These 
form the isometry group $I(M_4)=SO(3)\otimes U_4(1)$ of the Euclidean 
Taub-NUT space $M_4$. What is remarkable here is that the representation 
of $I(M_4)$ carried by $M_4$ mixes up linear transformations with non-linear 
ones involving the function $h$. 

In our opinion the study of this type of representation is important since it 
governs the transformation laws of the vectors and tensors under isometries 
that are the starting points in deriving conserved quantities 
through the Noether theorem.  For example, the isometry $(R,\lambda)$ 
transforms the contravariant components $v^\mu(x)$ of a four-dimensional 
vector field as 
\begin{eqnarray}
v^i(x)&\to &v^{\prime\,i}(x')=R^{i\,\cdot}_{\cdot\, j}v^j(x)\,,\\   
v^4(x)&\to &v^{\prime\,4}(x')=v^4(x)+v^i(x)\partial_i h(R,{\bf x})\,.\label{V4}
\end{eqnarray}
It is worth pointing out that from Eqs. (\ref{hA}) and (\ref{V4}) we can deduce 
that the quantity
\begin{equation}
v^{\prime\,4}(x')+{\bf A}({{\bf x}}')\cdot {{\bf v}}'(x')=
v^{4}(x)+{\bf A}({{\bf x}})\cdot {{\bf v}}(x)
\end{equation}
behaves as a scalar. These interesting properties of isometries could 
be better understood finding explicitly the analytical expression of 
the function $h$.   
 
\section{The induced representation}

The study of the function $h$ must combine the integration of the equations 
(\ref{hA}) with some algebraic properties resulted from the condition 
(\ref{cond}). The main point is to show that the transformation rule 
(\ref{ind}) of the fourth coordinate of $M_4$ is given by a representation 
of the isometry group induced by  one of its subgroups. We recall that the 
$SO(3)$ subgroup of $I(M_4)$ has three independent one-parameter subgroups, 
$SO_i(2)$, $i=1,2,3$, each one including rotations $R_i(\psi)$, of angles 
$\psi\in [0,2\pi)$ around the axis $i$. With this notation any rotation 
$R\in SO(3)$ in the usual Euler parametrization reads 
$R(\alpha,\beta,\gamma)=R_3(\alpha)R_2(\beta)R_3(\gamma)$.               

We start with the observation that because of the special form of gauge field 
(\ref{(tn)}) all the rotations of the subgroup $SO_3(2)$ satisfy the condition 
\begin{equation}\label{hatR}
\hat R {\bf A}({\bf x})= {\bf A}(\hat R {\bf x})\,,\quad \forall \, \hat R\in 
SO_3(2)\,. 
\end{equation}
Bearing in mind that, in addition, the gauge field does not depend on $x^4$ we 
say that the subgroup $H(M_4)=SO_3(2)\otimes U_4(1)\subset I(M_4)$ is by 
definition the {\em little group} associated to ${\bf A}$. 
In what follows we are interested to exploit the existence of the little group 
focusing on the rotations $\hat R\in SO_3(2)$. According to Eqs. (\ref{hA}) 
and (\ref{hatR}) it results that 
\begin{equation}\label{hath}
h(\hat R,{\bf x})\equiv \hat h(\hat R)  
\end{equation}
is point-independent being defined only on $SO_3(2)$. Then the condition 
(\ref{cond}) becomes 
\begin{equation}\label{abe}
\hat  h(\hat R\hat R')=\hat h(\hat R) + \hat h(\hat R')\,, \quad \forall 
~\hat R,\,\hat R' \in SO_3(2)
\end{equation}
which means that the set of values $\{\hat h(\hat R)\,|\, \hat R\in SO_3(2)\}$ 
form a one-di\-men\-sio\-nal representation of $SO_3(2)$ provided 
$\hat h(Id)=0$. This representation is non-trivial (with $\hat h(\hat R)\not=0$ 
when $\hat R\not=Id$) only if we assume that
\begin{equation}\label{hRa}
\hat h[R_3(\alpha)]= {\rm const.}\,  \alpha  \,.
\end{equation}
 
These properties suggest us to try to find the function $h$ using  
rotations in the Euler parametrization and the chart with spherical 
coordinates where the differential equations could be simpler since 
$h(R,{\bf x})=h(R,\theta,\phi)$ does not depend on the radial coordinate $r$. 
Then, according to Eqs. (\ref{hath}) and (\ref{abe}), we can write 
\begin{eqnarray}
h[R(\alpha,\beta,\gamma), {\bf x}]&=&\hat h[R_3(\alpha)]+h[R_2(\beta)
R_3(\gamma), {\bf x}] \nonumber\\
&=&h[R_2(\beta),R_3(\gamma){\bf x}]+\hat h[R_3(\alpha+\gamma)]
\end{eqnarray}    
pointing out that the central problem is to integrate the system (\ref{hA}) in 
spherical coordinates for the particular case of $R=R_2(\beta)$. 
Denoting  $h[R_2(\beta),{\bf x}]\equiv h(\beta, \theta,\phi)$, after a few 
manipulation we find that Eqs. (\ref{hA}) take the form
\begin{eqnarray}
\partial_{\theta}h(\beta,\theta,\phi)&=&-\mu\, \frac{\sin\phi\,\sin\beta} 
{1+\cos\theta\,\cos\beta-\sin\theta\,\cos\phi\,\sin\beta}\,,\label{cucu1}\\
\partial_{\phi}h(\beta,\theta,\phi)&=&\mu\, \frac{(1-\cos\theta)(1-
\cos\beta)-\sin\theta\,\cos\phi\,\sin\beta} 
{1+\cos\theta\,\cos\beta-\sin\theta\,\cos\phi\,\sin\beta}\,.\label{cucu2}
\end{eqnarray}
The integration of this system  gives $h(\beta,\theta,\psi)$ up to 
some arbitrary integration constants as it is shown in the Appendix. 
However, the definitive form of the function $h(R,\theta,\phi)$ 
can be easily found by using the technique of induced representations.

Let us consider the chart with spherical coordinates $(r,\theta,\phi,\chi)$ 
where $\theta$ and $\phi$ can be interpreted as the Euler angles of 
the rotation giving ${\bf x}=R(\theta,\phi,0){\bf x}_o$ from 
${\bf x}_o=(0,0,r)$. After an arbitrary rotation 
$R(\alpha,\beta,\gamma)\in I(M_4)$ we arrive to the chart with the new 
coordinates $(r, \theta', \phi', \chi')$ among them the first three 
are the spherical coordinates of the transformed vector  
\begin{eqnarray}
{{\bf x}\,}'&=&R(\phi',\theta',0){\bf x}_o=R(\alpha,\beta,\gamma){\bf x}
\nonumber\\
&=&R(\alpha,\beta,\gamma)R(\phi,\theta,0){\bf x}_o\,.\label{transx}
\end{eqnarray}
In addition, we assume that the fourth transformed spherical coordinate 
$\chi'$ is defined such that
\begin{equation}
R(\phi',\theta',\chi')=
R(\alpha,\beta,\gamma)R(\phi,\theta,0)
\end{equation}
from which we deduce
\begin{eqnarray}
R_3(\chi')&=&R^{-1}(\phi',\theta',0) R(\alpha,\beta,\gamma)
R(\phi, \theta, 0)\nonumber\\
&=&R^{-1}(\phi'-\alpha,\theta',0) R_2(\beta)R(\phi+\gamma, \theta, 0)\,.
\label{final}
\end{eqnarray}
Thus we obtain the transformation $R:\chi\to \chi'$ of a representation 
of the $SO(3)$ group {\em induced} by the subgroup $SO_3(2)$. The 
corresponding induced representation transforming the fourth Cartesian 
coordinate $x^4$ is defined by Eq. (\ref{ind}) where the function $h$ must be
\begin{equation}
h(R,\theta,\phi)=-\mu (\chi'+\phi'  -\phi)\,.
\end{equation}
The last step is to rewrite Eq. (\ref{final}) in terms of $SU(2)$ transformations 
corresponding to all the particular rotations involved therein. In this way it 
is not difficult to find the final result:
\begin{eqnarray}
h[R(\alpha,\beta,\gamma), \theta,\phi]&=&-\mu (\alpha+\gamma) \nonumber\\
&&-2\mu\, {\rm arctan}\left[\frac{\sin(\phi+\gamma)}
{\cot\frac{\theta}{2}\cot\frac{\beta}{2}-\cos(\phi+\gamma)}\right]\,.\label{FIN}
\end{eqnarray}
Now we can convince ourselves that this result is correct since 
for $\alpha=\gamma=0$ the function $h[R(0,\beta,0),\theta,\phi]=
h(\beta,\theta,\phi)$ is the desired solution of the system (\ref{cucu1}) 
- (\ref{cucu2}) calculated in Appendix. Moreover, if we put $\beta=\gamma=0$ 
we obtain that the point-independent functions (\ref{hRa}) are 
$h[R_3(\alpha),\theta,\phi]=\hat h[R_3(\alpha)]=-\mu\,\alpha$ (with 
const.=$-\mu$).

\section{The angular momentum}

In the quantum theory on the Euclidean Taub-NUT background the basic operators 
are introduced using the geometric quantization. Now when we know the closed 
form of the function $h$ we can calculate the generators of the natural 
representation of the group $I(M_4)$ using only group theoretical methods.

The natural representation is carried by the space of scalar wave functions 
$\Phi$ defined on $M_4$ which have the transformation rule 
$(R,\lambda):\Phi(x)\to \Phi'(x')=\Phi(x)$ under isometries 
$(R,\lambda)\in I(M_4)$. In this 
representation the generators are the (orbital) differential operators defined  
as usual. More precisely, denoting by $\xi$ a set of parameters of $I(M_4)$ and by 
 $x'(x,\xi)$ the transformed coordinates, one associates to  each parameter 
$\xi^a$ the Killing vector $k_a$ of components 
\begin{equation}\label{Kill}
k_a^{\mu}(x)=\left.\frac{\partial 
x^{\prime\, \mu}(x,\xi)}{\partial {\xi^a}}\right|_{\xi=0}\,,
\end{equation}
and the vector field
\begin{equation}
X_a=-ik_a^{\mu}(x)\partial_{\mu}
\end{equation}
that is just the generator of the natural representation corresponding to 
$\xi^a$.  In the case of our isometry group $I(M_4)$ we take the first three 
parameters, $\xi^i$, the Cayley-Klein parameters of the one-parameter subgroups 
$SO_i(2)$ and $\xi^4=\lambda$. Then we find that the generator  
of the $U_4(1)$ translations is the fourth component of the momentum operator,    
$P_4=-i\partial_4$. Furthermore, exploiting the form of $h$ we 
calculate the rotation generators in Cartesian coordinates. These are the 
components of the orbital angular momentum operator, 
\begin{eqnarray}     
L_1&=&-i(x^2\partial_3-x^3\partial_2)+i\mu\,\frac{x^1}{r+x^3}\partial_4\,\\ 
L_2&=&-i(x^3\partial_1-x^1\partial_3)+i\mu\,\frac{x^2}{r+x^3}\partial_4\,\\ 
L_3&=&-i(x^1\partial_2-x^2\partial_1)+i\mu\,\partial_4\,. 
\end{eqnarray}
The last terms arising from Eqs. (\ref{hxi}) are due to the existence of 
the function $h$ giving the induced representation discussed above. Thus we 
recover the known expression of the angular momentum operator 
\begin{equation}\label{(angmom)}
{\bf L}\,=\,{\bf x}\times{\bf P}-\mu\frac{{\bf x}}{r}P_4
\end{equation} 
where  ${\bf P}=-i({\mb \partial}-{\bf A}\partial_{4})$
is the three-dimensional (physical) momentum whose  components obey the 
commutation rules $[P_{i},P_{j}]=i\epsilon_{ijk}B_{k}P_{4}$ and 
$[P_{i},P_{4}]=0$. 

The scalar quantum mechanics  in the Taub-NUT geometry \cite{CFH} is based on 
the Schr\" odinger or Klein-Gordon equations  involving the  static operator    
\begin{equation}
\Delta=-\nabla_{\mu}g^{\mu\nu}\nabla_{\nu}
=V{{\bf P}\,}^{2}+\frac{1}{V}{P_{5}}^{2}
\end{equation}
which is either proportional with the Hamiltonian operator 
of the Schr\" odinger theory ($\Delta=2H$) or represents the static part of 
the Klein-Gordon operator \cite{CV5}. In both cases when an operator 
commutes with $\Delta$ we say that this is conserved. One can verify that 
the operators $L_i$ are  conserved  and satisfy the canonical commutation rules. 
Moreover, their commutators with the coordinates and the momentum operators 
are the usual ones. 

Other three important conserved operators are given by the three 
specific Killing tensors of the Taub-NUT geometry,  ${\bf k}_{\mu\nu}$  
\cite{CFH}. With their help one defines the vector operator which plays 
the same role as the Runge-Lenz vector in the usual quantum mechanical Kepler 
problem:
\begin{equation}
{\bf K}\,=-\frac{1}{2}\nabla_{\mu}{\bf k}^{\mu\nu}\nabla_{\nu}=
\frac{1}{2}\left({\bf P}\times {\bf L}-
{\bf L}\times {\bf P}\right)-
\mu \frac{{\bf x}}{r}
\left(\frac{1}{2}\Delta-\frac{1}{\mu^{2}}Q^{2}\right)
\end{equation}
with
\begin{equation}\label{Q}  
Q=-\mu P_4=-i\partial_{\chi}\,.
\end{equation}
This transforms as a vector under space rotations since its components obey
\begin{equation}
[L_{i},K_{j}]=i\epsilon_{ijk}K_{k}\,.
\end{equation}  
Moreover, the operators $K_i$ can be rescaled such that in association with 
the operators $L_i$ should constitute the basis generators of different 
dynamical algebras, $so(4)$, $so(3,1)$ or $e(3)$, corresponding to different 
domains of energy of the free quantum motion of scalar particles in the 
Taub-NUT background \cite{CFH}.    
In our opinion, all this machinery works because of the special form of the 
operators $L_i$ whose supplemental terms are produced by the induced 
representation discussed here. 
 
When one goes to the chart with spherical coordinate a special attention 
must be paid to the definition (\ref{4cp}) which shows that the fourth 
spherical coordinate $\chi$ is {\em translated} with $\phi$. This means that 
the whole algebra of observables in Cartesian coordinates must be transformed 
according to this translation into an {\em equivalent} algebra. 
To do this it is convenient to  define the translation operator in 
terms of the operator $Q$ (\ref{Q}):
\begin{equation}
U(\phi)=e^{i\phi Q} 
\end{equation}
which has to transform any operator $X$ defined in the Cartesian chart, 
giving the corresponding observable $\tilde X=U(\phi)XU^{\dagger}(\phi)$ in the 
spherical chart. In other words in the spherical charts we use a  
representation of the operator algebra different from that considered in 
Cartesian charts. Thus the components of the orbital angular momentum 
(\ref{(angmom)}) in the canonical basis 
(with $\tilde L_{\pm}=\tilde L_{1}\pm i\tilde L_{2}$) become
\begin{eqnarray}
\tilde L_{3}&=&-i\partial_{\phi}\,,\label{tl1}\\
\tilde L_{\pm}&=&e^{\pm i\phi}\left[\pm\,\partial_{\theta}+
i\left(\cot\theta\,\partial_{\phi}-
\frac{1}{\sin\theta}\,\partial_{\chi}\right)\right]\,.\label{tl2}
\end{eqnarray}
Many other operators including the Runge-Lenz vector will take new forms in 
this representation but preserving their commutation relations.   

In theories with spin the form of the conserved operators associated 
with isometries is, in addition, strongly dependent on the gauge one uses 
\cite{CART,ES}. 
The space $M_4$ allows a special Cartesian gauge in the chart with 
Cartesian coordinates \cite{BuCh} which leads to a total angular momentum 
having standard components $J_i=L_i+S_i$, with point-independent 
spin operators $S_i$ \cite{CV2}. The corresponding operators in the spherical 
chart are $\tilde J_i=\tilde L_i +S_i$ since $S_i$ commute with  
$U(\phi)$. We note that the Runge-Lenz operator in theories with spin half has 
specific spin terms as we have recently pointed out \cite{CV3,CV4}. It is 
interesting that the basic commutation relations of the algebra of observables 
are the same in the scalar theory as well as in theories with spin 
$\frac{1}{2}$ 
of Pauli or Dirac type \cite{CV4}.    
  
\section{New harmonics and spherical spinors}

It is natural that the unusual form of the $SO(3)$ generators (\ref{tl1}) and 
(\ref{tl2}) should lead to new spherical harmonics. These have to be the 
common eigenfunctions of complete set of commuting operators 
$\{\tilde{\bf L}^{2},\tilde L_{3},Q\}$  representing a basis of 
the Hilbert space of square integrable functions of $\phi$, $\theta$ and 
$\chi$ defined on the compact domain  $S^{2}\times D_{\chi}$. In 
Ref. \cite{CV1} we introduced the  $SO(3)\otimes U(1)$-harmonics, 
$Y^{q}_{l,m}$, 
which satisfy the eigenvalue problems
\begin{eqnarray}
\tilde{\bf L}^{2}Y_{l,m}^{q}&=&l(l+1)\,Y_{l,m}^{q}\,,\label{(lp)}\\
\tilde L_{3}Y_{l,m}^{q}&=&m\,Y_{l,m}^{q}\,,\label{(l3)}\\
QY_{l,m}^{q}&=&q\,Y_{l,m}^{q}\,,\label{(l0)}
\end{eqnarray}
and the orthonormalization condition 
\begin{eqnarray}
&&\left<Y_{l,m}^{q},Y_{l',m'}^{q'}\right>=
\int_{S^2}d(\cos\theta)d\phi\,\int_{0}^{4\pi}d\chi\,
{Y_{l,m}^{q}(\theta, \phi, \chi)}^{*}\,
Y_{l',m'}^{q'}(\theta, \phi, \chi)\nonumber\\
&&~~~~~~~~~~~~~~~~~~~~~~~~~~~~~=\delta_{l,l'}\delta_{m,m'}
\delta_{q,q'}\,,\label{(spy)}\,.\label{scpr}
\end{eqnarray}

The boundary conditions on $S^{2}\times D_{\chi}$ require $l$ and 
$m$ to be integer numbers while $q=0,\pm 1/2,\pm 1,...$ \cite{CFH} but, in 
general, $q$ can be any real number. 
Solving  Eqs. (\ref{(l3)}) and (\ref{(l0)}) we get
\begin{equation}\label{(Y)}
Y_{l,m}^{q}(\theta,\phi,\chi)=\frac{1}{4\pi}\, \Theta_{l,m}^{q}(cos\theta)
e^{im\phi}e^{iq\chi}
\end{equation}
where the function $\Theta_{l,m}^{q}$ calculated according to Eq. (\ref{(lp)}) 
must satisfy the normalization condition
\begin{equation}\label{(np)}
\int_{-1}^{1}d(\cos\theta)\left|\Theta_{l,m}^{q}(\cos\theta)\right|^{2}=2\,.
\end{equation}  
resulted from Eq. (\ref{scpr}). This problem has solutions for any values 
of the quantum numbers obeying 
\begin{equation}\label{(qml)}
|q|-1<|m|\le l
\end{equation} 
when one founds \cite{CV1}   
\begin{eqnarray}
\Theta_{l,m}^{q}(\cos\theta)&=&\frac{\sqrt{2l+1}}{2^{|m|}}\left[\frac{
(l-|m|)!\,(l+|m|)!}{
\Gamma(l-q+1)\Gamma(l+q+1)}\right]^{\frac{1}{2}}\label{(fin)}\label{(Tet)}\\
&&\times \left(1-\cos\theta\right)^{\frac{|m|-q}{2}}
\left(1+\cos\theta\right)^{\frac{|m|+q}{2}}\,
P_{l-|m|}^{(|m|-q,\,|m|+q)}(\cos\theta)\,.\nonumber
\end{eqnarray}
For $m=|m|$ the $SO(3)\otimes U(1)$ harmonics are given by (\ref{(Y)}) 
and (\ref{(fin)}) while for $m<0$ we have to use the obvious formula
\begin{equation}
Y_{l,-m}^{q}=(-1)^{m}\left(Y_{l,m}^{-q}\right)^{*}\,.
\end{equation} 
When the boundary conditions allow half-integer quantum numbers $l$ and $m$
then we say that the functions defined by Eqs. (\ref{(Y)}) and (\ref{(fin)}) 
(up to a suitable factor) represent $SU(2)\otimes U(1)$ harmonics.

Thus we have obtained a non-trivial generalization of the  spherical 
harmonics of the same kind as the spin-weighted spherical harmonics \cite{NP} 
or those studied in \cite{HARM}. Indeed, if $l$, $m$ and $q=m'$ are either 
integer or half-integer numbers then we have  
\begin{equation}
Y_{l,m}^{m'}(\theta,\phi,\chi)=\frac{\sqrt{2l+1}}{4\pi}D^{l}_{m,m'}
(\phi,\theta, \chi) 
\end{equation}
where $D_{m,m'}^{l}$ are the matrix elements of the irreducible representation 
of weight $l$ of the $SU(2)$ group corresponding to the rotation of Euler 
angles $(\phi, \theta, \chi)$. What is new here is that our harmonics  
are defined for any real number $q$. For this reason these are useful 
in solving some actual physical problems \cite{Gosh}. Notice that similar 
spherical harmonics were used recently in \cite{Mard} under a different name 
(ring-shaped harmonics)\footnote{The author seems to be not aware of 
the previous results of \cite{CV1}.}.     

With the help of these spherical harmonics we constructed the corresponding 
spherical spinors \cite{CV2} following the traditional method \cite{TH}. We 
defined the spherical spinors  $\Psi^{\pm}_{q,j,m_{j}}(\theta,\phi,\chi)$ 
as the common two-component eigenspinors of the eigenvalue problems  
\begin{eqnarray}
Q\,\Psi^{\pm}_{q,j,m_{j}}&=& q\,\Psi^{\pm}_{q,j,m_{j}}\,, \\
\tilde{\bf J}^{2}\,\Psi^{\pm}_{q,j,m_{j}}&=& j(j+1)\,\Psi^{\pm}_{q,j,m_{j}}
\,, \\
\tilde J_{3}\,\Psi^{\pm}_{q,j,m_{j}}&=& m_{j}\,\Psi^{\pm}_{q,j,m_{j}}\,, \\
(\sigma_ L+1_{2\times 2})\,\Psi^{\pm}_{q,j,m_{j}}&=& \pm(j+1/2)\,
\Psi^{\pm}_{q,j,m_{j}}\,,\label{kkk}  
\end{eqnarray}
with $\sigma_L=\sigma_i \tilde L_i$ where $\sigma_i$ are the Pauli matrices  
while $1_{2\times 2}$ is the $2\times 2$ identity matrix. These spinors are, in 
addition, eigenfunctions of 
$\tilde{\bf L}^{2}$ corresponding 
to the eigenvalues $l(l+1)$ with $l=j\pm\frac{1}{2}$. For $j=l+\frac{1}{2}
>|q|-\frac{1}{2}$ we have 
\cite{TH,CV1}  
\begin{equation}
\Psi^{+}_{q,j,m_{j}}=\frac{1}{\sqrt{2j}}\left(
\begin{array}{l}
\sqrt{j+m_{j}}\, Y^{q}_{j-\frac{1}{2},m_{j}-\frac{1}{2}}\\
\sqrt{j-m_{j}}\, Y^{q}_{j-\frac{1}{2},m_{j}+\frac{1}{2}}
\end{array}\right)
\end{equation}
while for $j=l-\frac{1}{2}>|q|-\frac{3}{2}$ we get
\begin{equation}
\Psi^{-}_{q,j,m_{j}}
=\frac{1}{\sqrt{2j+2}}\left(
\begin{array}{l}
\sqrt{j-m_{j}+1}\, Y^{q}_{j+\frac{1}{2},m_{j}-\frac{1}{2}}\\
-\sqrt{j+m_{j}+1}\, Y^{q}_{j+\frac{1}{2},m_{j}+\frac{1}{2}}
\end{array}\right)\,.
\end{equation}
These spherical spinors  are 
orthonormal since the $SO(3)\otimes U(1)$ harmonics are orthonormal with 
respect to the angular scalar product (\ref{scpr}).  

Finally, denoting $\sigma_r=\sigma_i x^i/r$ and using Eq.(\ref{kkk}) and the 
relation
\begin{equation}
 \left\{\sigma_r,\,
\sigma_{L}+1_{2\times 2}\right\}=2q\,1_{2\times 2}
\end{equation}
we find that    
\begin{equation}\label{pmmp}
\sigma_r \Psi^{\pm}_{q,j,m_{j}}=
\pm\lambda^{q}_{j}\,\Psi^{\pm}_{q,j,m_{j}}
+\sqrt{1-(\lambda^{q}_{j})^{2}}\,\Psi^{\mp}_{q,j,m_{j}}
\end{equation}
where $\lambda^{q}_{j}=q/(j+\frac{1}{2})$. Note that a 
similar result is obtained in Ref. \cite{HARM1}. By using this property
we have written down the particular solutions of the Dirac equation in the 
central charts of the Taub-NUT background \cite{CV2}.  

\section{Conclusions}

We showed here how transforms under rotations the fourth coordinate of the 
Euclidean Taub-NUT space pointing out that the transformation law is of 
a representation of the group $SO(3)$ induced by one of its $SO(2)$ subgroups. 
In this context the angular momentum operators get new specific terms that 
do not change their algebraic relations but give rise to new 
$SO(3)\otimes U(1)$ spherical harmonics and corresponding Pauli spinors.
To our knowledge the construction of these new objects 
of the Euclidean Taub-NUT geometry has not been well-described in 
the literature. 

In other respects, the four-dimensional Taub-NUT and Taub-NUT-AdS 
solutions of Einstein's equations play a central role in the 
construction of diverse and interesting M-theory configurations.
The present approach can be easily extended to higher-dimensional 
spacetimes solutions to the vacuum Einstein equations \cite{BB}. 
There are solutions with non-trivial topology describing spacetimes 
which are either locally AdS or locally asymptotically flat \cite{AC}.
Summing up, we believe that the extension of the present approach for 
the construction of the spherical harmonics to higher-dimensional 
spacetimes could be relevant within the context of M-theory and deserves 
further studies.

\subsection*{Acknowledgments}

We would like to thank L. Gy. Feher for suggesting to one of us (I.I.C.) a 
suitable approach to the problem presented here. 

\setcounter{equation}{0}\renewcommand{\theequation} 
{A.\arabic{equation}}

\section*{Appendix A\\The function $h(\beta,\theta,\phi)$}

Integrating separately the Eqs. (\ref{cucu1}) and (\ref{cucu2}) one obtains 
\cite{RG}
\begin{eqnarray}
h_{(1)}(\beta,\theta,\phi)&=&F(\beta,\phi)-2\mu\arctan 
\left[\frac{\tan\frac{\theta}{2}
\tan\frac{\beta}{2}-\cos\phi}{\sin\phi}\right]\,,\\
h_{(2)}(\beta,\theta,\phi)&=&\mu\phi+G(\beta, \theta)
-2\mu\arctan\left[ \frac{(\cos\theta+\cos\beta)\tan\frac{\phi}{2}}{1+
\cos(\theta+\beta)}\right]
\end{eqnarray}
where $F$ and $G$ are arbitrary functions. If we take $F=2\mu\phi-\mu\pi$ and 
$G=0$ then we find the function 
\begin{eqnarray}\label{FINB}
h(\beta,\theta,\phi)&=&
h_{(1)}(\beta,\theta,\phi)=
h_{(2)}(\beta,\theta,\phi)\nonumber\\
&=&-2\mu\arctan\left[\frac{\sin\phi}{\cot\frac{\theta}{2}
\cot\frac{\beta}{2}-\cos\phi}\right]
\end{eqnarray}
that is just the restriction of the function (\ref{FIN}) for $\alpha=\gamma=0$.

The terms of the angular momentum produced by $h$ have to be calculated 
according to Eqs. (\ref{Kill}) and (\ref{ind}) starting with
\begin{equation}
\left.
{\frac{\partial}{\partial\beta} h(\beta,\theta,\phi)}\right|_{\beta=0}=
-\mu \frac{\sin\theta\sin\phi}{1+\cos\theta}\,.
\end{equation}
Then, denoting $\xi^2=\beta$ for $\alpha=\gamma=0$ and $\xi^3=
\alpha$ for $\beta=\gamma=0$ and using a simple rotation of 
angle $\pi/2$ around the third axis one arrives at  
\begin{equation}\label{hxi}
\left.\frac{\partial h(R,{\bf x})}{\partial\xi^{1,2}}\right|_{\xi=0}=- 
\mu\frac{x^{1,2}}{r+x^3}\,,\quad
\left.\frac{\partial h(R,{\bf x})}{\partial\xi^3}\right|_{\xi=0}=- \mu
\end{equation}
where $R=R(\xi)$ is a rotation in the Cayley-Klein parametrization.

\end{document}